\documentclass[conference]{IEEEtran}
\IEEEoverridecommandlockouts
\usepackage{biblatex}
\usepackage{amsmath,amssymb,amsfonts, cleveref}
\usepackage{algorithmic}
\usepackage{graphicx}
\usepackage{textcomp}
\usepackage{xcolor}
\def\BibTeX{{\rm B\kern-.05em{\sc i\kern-.025em b}\kern-.08em
    T\kern-.1667em\lower.7ex\hbox{E}\kern-.125emX}}
\begin{document}

\title{Novelty Detection in Network Traffic:\\ Using Survival Analysis for Feature Identification}

\author{\IEEEauthorblockN{Taylor Bradley}
\IEEEauthorblockA{\textit{Johns Hopkins University}\\
Baltimore, MD \\
tbradl17@jh.edu}
\and
\IEEEauthorblockN{Elie Alhajjar}
\IEEEauthorblockA{\textit{Army Cyber Institute} \\
West Point, NY \\
elie.alhajjar@westpoint.edu}
\and
\IEEEauthorblockN{Nathaniel Bastian}
\IEEEauthorblockA{\textit{Army Cyber Institute} \\
West Point, NY \\
nathaniel.bastian@westpoint.edu}

}

\maketitle

\begin{abstract}
Intrusion Detection Systems are an important component of many organizations' cyber defense and resiliency strategies. However, one downside of these systems is their reliance on known attack signatures for detection of malicious network events. When it comes to unknown attack types and zero-day exploits, modern Intrusion Detection Systems often fall short. In this paper, we introduce an unconventional approach to identifying network traffic features that influence novelty detection based on survival analysis techniques. Specifically, we combine several Cox proportional hazards models and implement Kaplan-Meier estimates to predict the probability that a classifier identifies novelty after the injection of an unknown network attack at any given time. The proposed model is successful at pinpointing PSH Flag Count, ACK Flag Count, URG Flag Count, and Down/Up Ratio as the main features to impact novelty detection via  Random Forest, Bayesian Ridge, and Linear Support Vector Regression classifiers.
\end{abstract}


\section{Introduction}

In recent years, novelty detection, or the identification of unusual, unknown, and out-of-distribution data, has attracted a considerable amount of attention across various disciplines. This concept is deeply rooted in neurophysiology and seeks to model the way human brains learn from, categorize, and adapt to new stimuli.
 
In the cybersecurity domain, for example, intrusion detection systems (IDS) aim to identify if a network is under attack by detecting anomalies in network flows. Anomaly detection approaches label behavior that deviates from a normal model as anomalous, often making an implicit assumption that anomalies correspond to intrusive or problematic events. However, the conflation between anomalous events and intrusive events often accounts for higher than average false positive rates of anomaly-based IDS.
 
On the other hand, novelty detection is the task of classifying test data (at inference time) that differ in some respect from the data that are available during training of a machine learning classifier for some specific task. This may be seen as “one-class classification,” in which a model is constructed to describe normal training data. The novelty detection approach is typically used when the quantity of available abnormal data is insufficient to construct explicit models for non-normal classes \cite{Pimentel}.

 
In this paper, we present a novel approach to feature identification in support of novelty detection in the IDS setting using survival analysis techniques. We first train various machine learning classifiers to identify both benign traffic and known attack types using representative network flows of both traffic types. Then, we inject unknown attack type data and examine how the classifier identifies these attacks based on the features of the novel network flows. To this end, we use key concepts from survival analysis to analyze the features of the flow that are most influential in novelty detection. 

The outline of the paper is structured as follows. In Section \ref{RelWorks}, we discuss relevant literature on similar research in this domain. In Section \ref{Methods}, we discuss our methodology and give an overview of the performed experiments including how classifiers are chosen, trained, and injected with novel attack type data. In Section \ref{Results}, we discuss the resulting observations with emphasis on the application of survival analysis to assess feature identification and the interpretation of the different coefficients therein. In Section \ref{DisLim}, we give a brief description of the limitations of such an implementation, and we mention some potential directions for future work in Section \ref{Conc}.

\section{Related Works}\label{RelWorks}

Novelty detection, which is often also referred to as out-of-distribution (OOD) detection, is a task of identifying whether a test input is drawn far from the training distribution (in-distribution) or not. In general, the novelty detection problem aims to detect OOD samples where a detector is allowed access only to training data. Novelty detection is a classic yet essential problem in machine learning, with a broad range of applications, including medical diagnosis \cite{caruana2015intelligible}, fraud detection \cite{phua2010comprehensive}, and autonomous driving \cite{eykholt2018robust}.

Traditional machine learning techniques such as Support Vector Machine (SVM), Random Forest \cite{zhang2008random} and Adaptive Boosting \cite{hu2013online} have been used by researchers for constructing machine learning classifiers for the problem of network intrusion detection \cite{9006122}. However, the downside of these approaches, in general, is that they suffer from high false positive rate and low detection rate. These approaches are often not scalable to large datasets and the validation accuracy rarely scales as the size of the data increases.

Deep learning techniques are increasingly growing in popularity to address these problems. The CNN-BiLSTM model \cite{sinha2020efficient} presents an effective approach which stacks convolutional neural networks and bi-directional long short-term memory (LSTM) layers to learn and detect attacks in network flows. The DL-IDS model \cite{sun2020dl} is a hybrid approach to extract the spatial and temporal features of network flows and to provide a better intrusion detection system. The HAST-IDS model \cite{wang2017hast} learns the low-level spatial features of network flows using deep convolutional neural networks and then learns high-level temporal features using long short-term memory networks. Several additional models exist in the literature such as LuNet \cite{wu2019lunet} and DANTE \cite{ma2020dante}, and recent work shows success in using one-dimensional convolutional neural networks for detecting attacks in raw network packet captures rather than network flows (\textcite{mlnet}.

Survival analysis \cite{ref67} is a subfield of statistics that aims to model data where the outcome is the time until the occurrence of an event of interest. It was originally used in health data analysis and has since been employed in many applications, such as predicting student dropout time \cite{ameri2016survival} and insider threat detection \cite{alhajjar2021survival}. We extend these lines of effort here and devise yet another use case where survival analysis deems very useful for the task of analyzing a classifier's response when introduced to OOD data samples.

\section{Methodology}\label{Methods}

In this section, we describe the computational experiments performed for feature identification.

\subsection{Dataset Description}

In what follows, we adopt the Canadian Institute for Cybersecurity's Intrusion Detection Evaluation Dataset (CIC-IDS2017). CIC-IDS2017 is a comprehensive dataset containing both benign and up-to-date network attack data in the form of Wireshark PCAP files. The traffic is representative of naturalistic human behavior as it generates background traffic from the routine behaviors of 25 typical users interacting with common protocols such as HTTP, HTTPS, FTP, email, etc. \cite{Sharafaldin2018}. The data also covers a diverse set of cyber attack scenarios by recreating and executing network attack types from a list of common attack families using various open-source tools and code such as Patator, GoldenEye, Metasploit, etc. The attack profiles include \textbf{}
In what follows, we adopt the Canadian Institute for Cybersecurity's Intrusion Detection Evaluation Dataset (CIC-IDS2017). CIC-IDS2017 is a comprehensive dataset containing both benign and up-to-date network attack data in the form of Wireshark PCAP files. The traffic is representative of naturalistic human behavior as it generates background traffic from the routine behaviors of 25 typical users interacting with common protocols such as HTTP, HTTPS, FTP, email, etc. \cite{Sharafaldin2018}. The data also covers a diverse set of cyber attack scenarios by recreating and executing network attack types from a list of common attack families using various open-source tools and code such as Patator, GoldenEye, Metasploit, etc. The attack profiles include Brute Force Attack, Heartbleed Attack, Botnet Attack, DoS Attack, DDoS Attack, Web Attack and Infiltration Attacks.

Each of these broad attack families also consists of specific attack types using various tools and techniques. For example, the web attack class contains three individual attack types: brute force web attack, cross-site scripting, and SQL injection. For the experiments below, we focus on these individual attack types to get a more consistent training distribution.

The dataset also includes network traffic analysis data in the form of labeled flows extracted using the CICFlowMeter, a network traffic flow generation tool. CICFlowMeter analyzes raw network traffic based on time stamp, source/destination IP's, source/destination ports, protocols, and attack type to generate 84 network traffic features \cite{Lashkari2018}. These features include details about a particular communication such as flow duration,  packet size statistics, various flag counts, flow length metrics, number of packets/bytes sent in a particular time window, etc. For our experiments, we chose to focus strictly on the network flow components of the CIC-IDS dataset since these features are a representative summary of a particular network communication. This allowed us to properly analyze the features of a network communication that are most influential in novelty detection. 

In this study, we train multiple machine learning classification models to detect instances of novelty in network flows. Specifically, we make use of the Random Forest Regressor, Bayesian Ridge, and Linear Support Vector Regression. We first split the CIC-IDS data up by individual attack type and tool such as DoS Hulk, FTP-Patator, SQL Injection, etc. Next, we train each of our three models using flows from both benign traffic and one single attack type. After training, we test the accuracy of the model's predictions to ensure that it has been adequately trained to distinguish between benign traffic and the corresponding attack type. We then inject unknown, or novel, attack type flows and assess how the model classifies this data. Since we use regression models, the continuous output variable represents how close the classifier considers this data is to each of the known classes. For example, if a predicted value of 0.45 is returned, this means that the classifier believes this flow to be ``benign" with very low confidence since it is a closer match to 0 than 1 but lies around the middle point. We define this classifier confusion within a given interval as ``novelty detection". 

\subsection{Survival Analysis}

Survival analysis is a subfield of statistics that focuses on estimating the time until, and frequency of, a particular event of interest. One of the most common applications is the context of medical studies, in which survival analysis techniques are used to predict a patient's survival time based on various features such as sex, age, and disease stage. However, the use of these statistical models stretches far beyond medicine. Survival analysis has particularly relevant applications in various technology disciplines for purposes such as estimating the time until a system failure, insider threat instance, or data breach occurs \cite{alhajjar2021survival}. These techniques also deal with the issue of censoring, which occurs when the event of interest is never observed due to withdrawal from the study or the study ending prior to that event's potential occurrence. In any case, our observation, $Y$, is either survival time, $T$, or censoring time, $C$. Using the status indicator $\delta$, if $\delta=1$, we observe true survival time, otherwise if $\delta=0$, we observe censoring time \cite{JamesBook}.

\begin{equation}Y = \text{min}(T,C),\end{equation}
\begin{equation}\delta = \begin{cases}1 & \text{if } T\leq C\\ 0 & \text{if } T>C.\end{cases}\end{equation}

There are three main functions used in survival analysis: 

\begin{itemize}
    \item Survival function: $S(t) = Pr(T \geq t)$
    \item Cumulative death distribution function: $F(t) = 1 - S(t)$ 
    \item Death density function: $f(t) = F'(t)$ or $f(t) = \frac{dF}{dt}$
    \item Hazard function: $\frac{f(t)}{S(t)}$
\end{itemize}

\subsubsection{Cox Proportional Hazards Model. }

The Cox proportional hazards model is commonly used to investigate the association between survival time and one or more predictor variables. Since this model requires no knowledge of the underlying distribution, the baseline hazard functions of individuals are assumed to be the same. For each data point in the Cox model, the hazard function can be defined by:

\begin{equation} \label{eq1}
h(t) = h_0(t)e^{\beta.X_i},  
\end{equation}

\noindent where $h_0(t)$ represents the baseline hazard function, $X_i$ is the feature vector (where $i$ runs over the size of the dataset), $\beta$ is the corresponding coefficient vector, and $\beta .X_i$ is the vector scalar product. In this equation, the baseline hazard function can be an arbitrary non-negative function since the baseline hazard is assumed to be the same for every data point. From this equation, it follows that the survival function can be computed as: 

\begin{equation}
    S(t) = S_0(t)e^{\beta.X_i}.
\end{equation}

Cox model parameters are estimated by maximizing the partial likelihood with respect to $\beta$. This partial likelihood is simply a product of the probabilities over all uncensored observations and can be constructed as:

\begin{equation}
 PL(\beta) = \prod_{\delta_i=1}\frac{e^{\beta.X_i}}{\sum_{t_j\geq t_i}e^{\beta.X_j}}.  
\end{equation}

This calculation is valid regardless of the true baseline hazard value, making this a very flexible and reliable model. In addition to coefficient estimates, we can also use this model to obtain other valuable model outputs such as associated p-values and confidence intervals \cite{JamesBook}. 

\subsubsection{Kaplan-Meier Estimates. }

The Kaplan-Meier curve is a visual representation of the survival function that represents the probability of survival after time $t$. This estimate is based on the quantity of event occurrences in the actual length of observed time. While Kaplan-Meier estimates offer a simplistic way to estimate the survival function, this estimate becomes slightly more complex in the presence of censored data. To overcome this, we let $T_1 < T_2 < ... < T_k$ represent a set of ordered event times among $N$ non-censored instances, where $k \leq N$ meaning that there are $N-k$ censored times. For each $i = 1,2,...,k$, we let $d_i$ represent the number of actual observed events in the time $T_i$, and $r_i$ represent the number of instances whose actual survival time $T$ or censored time $C$ is greater than or equal to $T_i$. We let $c_{i-1}$ represent the number of censored instances that occurred between the time period $T_{i-1}$ and $T_i$ which allows us relate the terms with the recursion $r_i = r_{i-1} - d_{i-1} - c_{i-1}$.

This setting allows us to overcome the issue of censoring by allowing the conditional probability of surviving beyond time $T_i$ to be defined as:

\begin{equation} \label{eqn: 7}
   Pr(T_i) = \frac{r_i - d_i}{r_i} = 1-\frac{d_i}{r_i}. 
\end{equation}

The conditional property represented in Equation 7 leads to the Kaplan-Meier estimator of the survival curve:

\begin{equation} \label{eq:8}
 \widehat{S}(t) = \prod_{T_i < t} Pr(T_i) = \prod_{T_i < t} (1-\frac{d_i}{r_i}).  
\end{equation}

\subsection{Experimental Design}

We seek to investigate the features that would be of most influence to a novelty detection agent. From here on, we call $D_{pre}$ the pre-novelty distribution, which consists of flows from a single known attack type as well as benign traffic. Similarly, we call $D_{post}$ the post-novelty distribution, which consists of flows from a new attack type \cite{Pinto2022}. 

For the proposed experiments, we set an arbitrary prediction threshold of 40-60\%, or 0.40 to 0.60, to indicate that a classifier has detected novelty. In this case, we define the ``death", or event of interest, as a classifier's prediction of injected $D_{post}$ data in that range. After training each classifier on flows from $D_{pre}$, we assess its predictions of unknown attack data. To do this, we first create a sequence of novel network activity by selecting a random sample of 100 network flows of a new attack type from the $D_{post}$ data. Each of these sequences is equivalent to one ``patient" in the context of survival analysis. Next, we create a dataset of 500 sequences, each containing 100 $D_{post}$ network flows, and feed each sequence to the classifier individually. Each flow's value is then predicted by the classifier as a decimal number between $0$ and $1$. If the predicted label falls within the defined range, we consider that sequence ``dead", meaning that the classifier at hand detected novelty as defined above. If it falls outside of that range on either side, the classifier moves on to the next flow in the sequence. 

While we understand that misclassification of novel, malicious traffic as benign has greater consequences than the reverse case where a unknown attack is classified as a known attack, examining cases of misclassification is not the objective of the current work. Instead, we are analyzing cases of classifier confusion when faced with OOD samples. Survival time, or the index in the sequence at which the novelty was detected, for each of the 500 sequences is recorded. If novelty is not detected in any of the 100 flows, the instance is considered censored and survival time is set to be the length of the sequence. We repeat 10 iterations of this experimental procedure for each combination of $D_{pre}$ and $D_{post}$ attack data to create the corresponding Kaplan-Meier curves and Cox coefficient estimates.

\begin{table}[thb]
\centering
\begin{center}
\resizebox{\columnwidth}{!}{
\begin{tabular}{ |c|c|c| } 
\hline
Combination & Training Data Types & Novel Attack Type \\
\hline
1 & Benign, DoS Hulk & Brute-Force Web \\ 
2 & Benign, DoS Hulk & FTP-Patator \\ 
3 & Benign, Portscan & Web XSS \\ 
4 & Benign, Portscan & DoS GoldenEye \\ 
\hline
\end{tabular}}
\end{center}
\caption{Data combinations used for training and injection.}
\label{fig: table1a}       
\end{table}

To analyze the features in the $D_{post}$ data that contributed to classifier confusion in each experimental iteration, we compare the feature values of the specific flow that was classified in novelty range to the mean of the corresponding feature values in the $D_{pre}$ data. This allows us to examine how heavily a feature's distance from the known distribution influences the confusion of the classifier. 
\begin{equation}
x_i = [f_1^i, f_2^i, f_3^i, ..., f_{84}^i],
\end{equation}
\begin{equation}
\overline{x_{pre}} =  [\bar{f_1}, \bar{f_2}, \bar{f_3}, ..., \bar{f_{84}}],
\end{equation}
\begin{equation} 
y = |x_i - \overline{x_{pre}}|.
\end{equation} 

Here, $x_i$ represents the feature vector of the detected novel flow, $\overline{x_{pre}}$ is the vector of the mean feature values for the entirety of the $D_{pre}$ dataset, $f_i$ represents the flow's 84 individual feature values, and $y$ represents the absolute difference between the feature value of the detected novel flow, to the mean of the corresponding feature value in the $D_{pre}$ data.

Next, we analyze the results of this experiment using both the Cox model and Kaplan-Meier estimate. To generate the Kaplan-Meier curves, we use the $D_{pre}$ flow indices to represent time. We divide the $D_{pre}$ flow indices into intervals, $t_0 < t_1 < ... < t_m$, where $t_0$ and $t_m$ are the starting and ending indices, respectively. In the terminology of Section 2, we define $d_j$ as the number of sequences in which novelty is detected by the classifier at index $t_j$ and $n_j$ as the number of sequences where novelty is not yet detected at index $t_j$. Using Equations 7 and 8, we can estimate the survival function as:

\begin{equation}
    S(i_{j+1}) = S(i_j) * (1-\frac{d_j}{n_j}).
\end{equation}

To apply the Cox proportional hazards model, we label each of the 500 sequences in the experiment as having detected novelty or not using binary indicators, 1 or 0, respectively. We use the 84 features of the $D_{post}$ flows for identification of risk factors associated with novelty detection. After maximizing the partial likelihood in Equation 5, we can set the derivative of that equation with respect to $\beta$ equal to zero. This allows us to estimate the coefficients for each feature and, as a result, the baseline hazard function. 

\begin{equation}
    ln(\frac{h(t)}{h_0(t)}) = \beta.X_i = \beta_1.X_{i1} + \beta_2.X_{i2} + ... + \beta_{84}.X_{i84}.
\end{equation}

\section{Results}\label{Results}

The mean resulting Cox proportional hazard coefficient estimates, $\beta$, from each of the 10 trials for each classifier and attack combination are shown in \Cref{fig: table1}. These results shed light on which network flow characteristics have the greatest impact on novelty detection for each classifier and attack combination, since coefficient estimates, $\beta$, measure the impact of each feature on the probability of novelty detection. For interpretability, we also compute the hazard ratios ($HR$) for each feature by exponentiating the parameter estimates ($\beta$) using the technique $e^\beta$. For any parameter estimate, if $HR<1$, the predictor is said to be protective, or associated with improved survival. In this case, this means that the individual flow feature is less likely to confuse the classifier and instead leads to classification in one of the two known categories. Conversely, if $HR>1$, the predictor is associated with increased risk, and therefore decreased survival. In the context of network systems, this means that the individual flow feature is more likely to confuse the classifier and lead to novelty detection. 

\begin{table}[thb]
\centering
\begin{center}
\resizebox{\columnwidth}{!}{
\begin{tabular}{ |c|c|c|c| } 
\hline
Classifier & Feature & $\beta$ & HR \\
\hline
Random Forest & PSH Flag Count & 0.157 & 1.170\\ 
& ACK Flag Count & -0.580 & 0.560\\ 
& URG Flag Count & -0.105 & 0.900\\
& Down/Up Ratio & 1.506 & 4.509\\
\hline
Bayesian Ridge & PSH Flag Count & 31.401 & $4.3x10^{13}$\\ 
& ACK Flag Count & -128.528 & $1.5x10^{-56}$\\ 
& URG Flag Count & -23.333 & $7.4x10^{-11}$\\
& Down/Up Ratio & 8.811 & $6.7x10^{3}$\\
\hline
Linear SVR & PSH Flag Count & 0.194 & 1.121\\ 
& ACK Flag Count & -0.813 & 0.444\\ 
& URG Flag Count & -0.144 & 0.866\\
& Down/Up Ratio & 0.682 & 1.978\\
\hline
\end{tabular}}
\end{center}
\caption{Cox model results for classifiers trained on DoS Hulk data and injected with Brute-Force web attack flows. }
\label{fig: table1}       
\end{table}

\begin{figure}
\centering
\includegraphics[width=240pt]{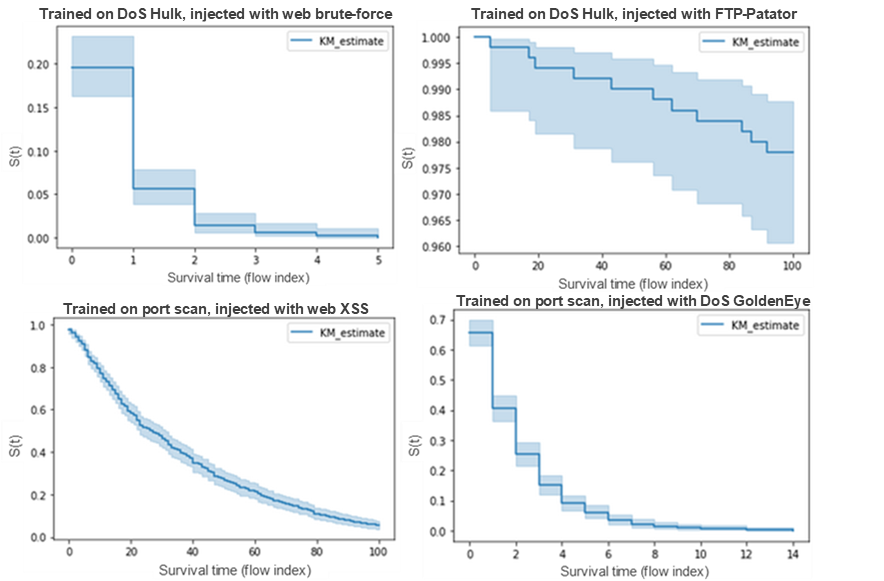}
\caption{Kaplan-Meier curves for Random Forest classifier with each attack combination.}
\label{fig:rf}
\end{figure}

The resulting Kaplan-Meier curves for each attack combination and classifier allow us to estimate the probability of the classifier detecting novelty at sequence index or time $t$ and are shown in \Cref{fig:rf}.

\section{Discussion and Limitations}\label{DisLim}

The results for the Cox model shown in \Cref{fig: table1} highlight the impact of the network features on the likelihood of novelty detection for each classifier. After applying the Cox model, we found four consistently non-zero coefficients across all classifiers and attack types which allowed us to narrow down the most influential features. For example, holding all other factors constant, there is a 17\% increase in the likelihood of novelty detection relative to a one unit increase in the number of packets with a PSH flag contained in the flow. Conversely, there is a 48\% decrease in the likelihood of novelty detection per one unit increase in the flow's URG flag count. 

Given that the signs of the coefficients remain the same regardless of the classifier type, we summarize the findings as follow:
\begin{enumerate}
    \item Higher PSH flag counts are positively correlated with novelty detection for this attack combination. One way to justify such a relationship is that PSH flags are used to indicate that data should be pushed to the receiving host immediately. They are typically used in TCP connections to inform the receiver that the sender has no data left to send. Because of their purpose, they are typically not used in large quantities in DoS attacks, therefore they are unfamiliar to the classifier and more likely to indicate a novel network attack.  
    \item Higher ACK flag counts are negatively correlated with novelty detection. This can be explained by the fact that this DoS attack floods the target system with an abundance of connection initiation requests so it is probable that the target sends many ACK replies before becoming overwhelmed and losing availability. Since DoS data was used for training, the classifier is very familiar with ACK flags. When it recognizes higher ACK flag counts in the injected brute-force web attack data, it is more likely to misclassify it as a known instance. 
    \item Similarly to ACK flags, the higher the URG flag count, the less likely the flow is to confuse the classifier and indicate novelty. Since URG flags are commonly used in DoS attacks to quickly overwhelm the server by indicating that the requests should be prioritized, this again means that the classifier is familiar with these types of flags, so higher counts would lead to misclassification.
    \item Higher file download/upload ratios are positively correlated with novelty detection. This is due to the fact that the DoS Hulk attack did not include downloads or uploads of any kind. Therefore, these types of packets are likely deemed unknown, unfamiliar, and ultimately correspond to a novel attack type. 
\end{enumerate}

Using the Kaplan-Meier curves shown in \Cref{fig:rf}, we see each classifier's success at detecting novelty with each attack combination presented in \Cref{fig: table1a}. To better understand this observation, we examine Figure 1. From these curves, we can draw a few conclusions about the Random Forest classifier:
\begin{enumerate}
    \item The Random Forest classifier detects novelty the quickest for attack combination 1, with 100\% detection, or ``death", by flow index 5. This included 80\% detection at index 0.
    \item For attack combination 2, the Random Forest classifier only detects novelty in approximately 5\% of the flows by index 100, making the other 95\% of these instances censored.
    \item For attack combination 3, the classifier detects approximately 98\% novelty by index 100.
    \item Lastly, for attack combination 4, the classifier reaches 100\% detection by index 14 which includes approximately 35\% detection at index 0.
\end{enumerate}

While these results shed light on the various features of network flows that can assist in novelty detection, there are a few limitations to this study. Firstly, our experiments do not directly account for misclassifications of novel attack types as benign. Since we focus specifically on novelty detection, we saved these results for future analysis in an upcoming paper. Secondly, we found that many other types of classifiers actually do not detect novelty at all with any combination of attack types. This could lend itself to future work in investigation of the various classifier decision making processes that lead to the classification of these novel attacks as known. Lastly, our experiments do not represent the entirety of the CIC-IDS dataset. This is because many attack types are severely underrepresented which may skew the results and draw statistically insignificant conclusions. For example, there are only 11 labeled Heartbleed flows in the entirety of CIC-IDS dataset, making it difficult to train or test on this particular attack type.  

\section{Conclusion}\label{Conc}

With the growing importance of machine learning for use in the cybersecurity world, the concept of novelty detection is more important than ever. Given the underexplored nature of this issue, it is important to examine how commonly used classifiers may be adapted to detect novelty. Hence, understanding and gaining insight into the features that may help IDS increase their effectiveness against zero-day attacks is an essential task. In this paper, we introduced a survival analysis based approach to discovering the most influential network flow features in various classifiers' novelty detection ability. We used the CIC-IDS dataset to train classifiers, inject unknown attack types, assess the classifiers ability to detect novelty, and as a result, the features that play the greatest role in that detection. Our results showed that PSH, URG and ACK flag counts, as well as download/upload ratios, play a role in novelty detection across all classifiers and attack combinations. 

There remain many challenges in the field of IDS development and resilience. First, there are not many publicly available and easily accessible solutions to novelty detection, making them much more difficult to implement. Second, there is a vast misunderstanding of the difference between novelty detection and anomaly detection and their implementations. This often causes these techniques to be overlooked by IDS developers. It is impractical to train any given system on every single attack that has ever happened, along with every normal behavior. In addition to this, balancing the distributions of the many classes in a training dataset poses a significant challenge moving forward. Lastly, there is a lack of realistic datasets that represent up-to-date attack type data. This makes it impossible to train robust IDS, as these systems rely on known attack signatures that must be continually updated to detect the latest attacks. Moreover, it severely limits the test bed for novelty detection agents in this domain.

\section*{Acknowledgements}
This work is supported in part by the National Security Agency Laboratory for Advanced Cybersecurity Research under Interagency Agreement No. USMA 21035 and the U.S. Army Combat Capabilities Development Command (DEVCOM) Army Research Laboratory under Support Agreement No. USMA 21050. The views expressed in this paper are those of the authors and do not reflect the official policy or position of the United States Military Academy, the United States Army, the Department of Defense, or the United States Government.

\printbibliography

\end{document}